\documentclass[english,fleqn,reviewcopy]{elsart}
\usepackage[T1]{fontenc}
\usepackage[latin1]{inputenc}
\usepackage{babel}
\usepackage{graphics}
\usepackage{amsmath}
\usepackage{amssymb}

\makeatletter
\renewcommand{\(}{$}

\newcommand{\cal}{\mathcal}
\makeatother
\begin{document}
\newcommand{\C}{{\mathcal{C}}} \newcommand{\V}{{\mathcal{V}}}

\begin{frontmatter}
{\raggedleft hep-ph/0211398, MSUHEP-21125, SMU-02/05\par}

\title{Diphoton production in gluon fusion at small\\ transverse momentum}

\author[Pavel]{P. M. Nadolsky} and
\author[Carl]{C. R. Schmidt}

\address[Pavel]{Department of Physics, Southern Methodist University, Dallas,
TX 75275, U.S.A.}
\address[Carl]{Department of Physics \& Astronomy, Michigan State University,
East Lansing, MI 48824, U.S.A.}

\begin{abstract}
We discuss the production of photon pairs in gluon-gluon scattering
in the context of the position-space resummation formalism at small
transverse momentum.  We derive the
remaining unknown coefficients that arise at \({\mathcal{O}}( \alpha _{S}) $,
as well as the remaining \({\mathcal{O}}(\alpha_{S}^{2})$ coefficient 
that occurs
in the Sudakov factor.  
We comment on the impact of
these coefficients on the normalization and shape  of the
resummed transverse momentum distribution
of photon pairs, which comprise an important background to
Higgs boson production at the LHC.
\end{abstract}
\end{frontmatter}

The Standard Model production of photon pairs with a large invariant mass
plays a vital role in physics studies at the Large Hadron Collider (LHC).
It provides a large background to the production of Higgs bosons, where the
Higgs boson subsequently decays in the diphoton channel 
($ pp\rightarrow HX\rightarrow \gamma \gamma X $).
Despite the small branching ratio of the Higgs boson to two photons, this
mode is the most important one for $ M_{H}\lesssim140  $ GeV, due to the
narrow width of the Higgs boson and the fine mass resolution of photon pairs
in the LHC detectors \cite{LHCproposals}, which allow a Higgs boson peak
to be found above the continuum background. The efficient discrimination
of Higgs boson events from the background relies on the accurate knowledge
of the kinematic distributions of both signal and background. In a recent
paper \cite{BalazsNadolskySchmidtYuan}, we and our collaborators discussed
the diphoton background and calculated the transverse momentum distribution
of the photon pairs in the framework of the Collins-Soper-Sterman (CSS) 
resummation
formalism \cite{CS,CSS}. This resummation is necessary to handle correctly
the large effects of soft and collinear QCD radiation at diphoton transverse
momenta $ Q_{T} $ of about $ M_{H}/2 $ or less.

In Ref.~\cite{BalazsNadolskySchmidtYuan}, significant attention was paid
to the production of photon pairs in gluon-gluon fusion 
$ gg\rightarrow \gamma \gamma X $.
This subprocess first arises at $ {\mathcal{O}}(\alpha _{S}^{2}) $ in
the perturbative expansion in the QCD coupling. Thus, it is formally of a
higher order than the quark annihilation subprocess 
$ q\bar{q}\rightarrow \gamma \gamma X $,
which enters at $ {\mathcal{O}}(\alpha _{S}^{0}) $. Despite the extra
factors of $ \alpha _{S} $, the two contributions are comparable numerically,
because of the large gluon luminosity in the relevant mass range at the LHC.
Furthermore, the lowest order (LO) $ gg\rightarrow \gamma \gamma  $ 
contribution
occurs through a one loop box diagram, which is infrared finite and is not
related through factorization to the $ {\mathcal{O}}(\alpha ^{0}_{S}) $
and $ {\mathcal{O}}(\alpha ^{1}_{S}) $ diagrams in the quark annihilation
channel. Therefore, it can be treated as the LO diagram of an independent
perturbative contribution to diphoton production.

Recently, the complete next-to-leading (NLO) cross section for the gluon
fusion subprocess has been calculated \cite{BernDixonSchmidt}. That calculation
utilized the cross sections for the 
$ {\mathcal{O}}(\alpha ^{2}\alpha _{S}^{3}) $
real emission subprocess $ gg\rightarrow \gamma \gamma g $ 
\cite{BalazsNadolskySchmidtYuan,deFlorianKunszt}
and the recently-computed two-loop virtual corrections to the 
$ {\mathcal{O}}(\alpha _{S}^{2}) $
box diagram \cite{BernDeFreitasDixon}. In this paper, we use the results
of the above publications to derive all the NLO coefficients in the resummed
cross section and the remaining unknown NNLO coefficient in the perturbative
Sudakov factor.

In the CSS formalism, the gluon-fusion cross section at small transverse
momentum can be expressed as a Fourier-Bessel transform of a form factor,
\break
$ \widetilde{W}(b,Q,v,x_{A},x_{B}) $, in terms of the impact parameter
$ b $: 
\begin{equation}
\left. \frac{d\sigma (gg\rightarrow \gamma \gamma X)}{dQ^{2}dydvdQ_{T}^{2}}
\right| _{Q_{T}\rightarrow 0}\approx \int \frac{d^{2}b}{(2\pi )^{2}}
e^{i\vec{Q}_{T}\cdot \vec{b}}\widetilde{W}(b,Q,v,x_{A},x_{B}).
\end{equation}
 The perturbative part of $ \widetilde{W}(b,Q,v,x_{A},x_{B}) $ can be
written as
\begin{eqnarray}
&& \widetilde{W}(b,Q,v,x_{A},x_{B})=\frac{\sigma _{0}
(\alpha _{S}(\mu _{0}))}{S} \nonumber \\
 &  & \times \exp \left\{ -\int _{C_{1}^{2}/b^{2}}^{C^{2}_{2}Q^{2}}
\frac{d\bar{\mu }^{2}}{\bar{\mu }^{2}}
\Bigl [{\mathcal{A}}(\alpha _{S}(\bar{\mu }))
\ln \frac{C_{2}^{2}Q^{2}}{\bar{\mu }^{2}}+
{\mathcal{B}}(\alpha _{S}(\bar{\mu }))\Bigr ]\right\} \nonumber \\
 &  & \times \sum _{a,b=g,\Sigma }
\left[ \C _{g/a}\otimes 
f_{a}\right] (x_{A}, b;\mu _{F}, \alpha_S(\mu_F))\left[ \C _{g/b}
\otimes f_{b}\right] (x_{B}, b;\mu _{F},\alpha_S(\mu_F)).
\end{eqnarray}
 Here $ Q,\, y, $ and $ Q_{T} $ are the invariant mass, rapidity and
transverse momentum of the photon pair, respectively; $ S $ is the square
of the $ pp $ center-of-mass (c.m.) energy; 
$ v\equiv (1-\nolinebreak \cos \theta ^{*})/2 $,
where $ \theta ^{*} $ is the polar angle of one of the photons in the
$ \gamma \gamma  $ c.m. frame; and $ x_{A,B}\equiv Qe^{\pm y}/\sqrt{S} $.
The momentum scale at which the QCD coupling $ \alpha _{S} $ is evaluated
is shown explicitly in each of the terms. The convolution is defined in the
conventional manner, 
\begin{equation}
\left[ f\otimes g\right] (x)=\int _{x}^{1}
\frac{d\xi }{\xi }f(\xi )g\left( x/\xi \right) \, \, .
\end{equation}
 The summation over the indices $ a $ and $ b $ goes over the gluon
parton distribution function (PDF) $ f_{g}(x,\mu _{F}) $ and the quark
singlet PDF $ f_{\Sigma }(x,\mu _{F}) $, which are evaluated at a momentum
scale $ \mu _{F}. $ The parameters $ C_{1} $ and $ C_{2} $ in the
Sudakov term are constants of order unity. 
In general the scales $\mu_0$ and $\mu_F$ should be of order $Q$ and
$1/b$, respectively, so as not to introduce large logarithms in eq.~(2).
For the process 
$ gg\rightarrow \gamma \gamma X $,
the normalization factor is 
\begin{equation}
\label{normfact}
\sigma _{0}\, \, =\, \, \frac{\alpha ^{2}\alpha _{S}^{2}(\mu _{0})
\left( \sum q_{i}^{2}\right) ^{2}\sum |M^{(1)}|^{2}}{64\pi Q^{2}}\, \, ,
\end{equation}
 where $ q_{i} $ are the charges (in units of $ e $) of the quarks
that run in the box loop, and the second summation is over the helicities
$ \lambda _{1},\lambda _{2},\lambda _{3},\lambda _{4} $ of the gluons
and photons. The LO helicity amplitudes $ M^{(1)}\equiv 
M^{(1)}_{\lambda _{1}\lambda _{2}\lambda _{3}\lambda _{4}} $
are given in Eq.~(3.15) of Ref.~\cite{BernDeFreitasDixon}. They can be
expressed as functions of $ v=(1-\cos \theta ^{*})/2=-\hat{t}/\hat{s} $,
where $ \hat{t} $ and $ \hat{s} $ are Mandelstam variables of the LO
2-to-2 process.

The functions $ {\mathcal{A}}(\alpha _{S}),\, {\mathcal{B}}(\alpha _{S}), $
and $ \C _{g/a}(x, b; \mu_F, \alpha _{S})$ can be expanded as 
a perturbation series
in $ \alpha _{S} $: 
$ {\mathcal{A}}(\alpha _{S})=\sum _{n=1}^{\infty }
\left( \alpha _{S}/\pi \right) ^{n}{\mathcal{A}}^{(n)},
\,\, {\mathcal{B}}(\alpha _{S})=\sum _{n=1}^{\infty }
\left( \alpha _{S}/\pi \right) ^{n}{\mathcal{B}}^{(n)} $,
and
${\C}_{g/a}(x, b; \mu_F, \alpha _{S}) = \delta _{ag}\delta (1-x)+
\sum _{n=1}^{\infty }
\left( {\alpha _{S}}/{\pi }\right) ^{n}{\C}^{(n)}_{g/a}(x)
$. 
For brevity, we suppress the explicit dependence 
of $C^{(n)}_{g/a}(x)$ on $b$ and $\mu_F$.
The coefficients  ${\mathcal{A}}^{(1)} $, $ {\mathcal{B}}^{(1)} $, and  
$ {\mathcal{A}}^{(2)} $ 
in the Sudakov factor have been known
for some time \cite{Catani:vd}:
\begin{eqnarray}
{\mathcal{A}}^{(1)} & = & N_{c}\,,\qquad
{\mathcal{B}}^{(1)}=-\beta _{0}-2N_{c}\ln \frac{b_{0}C_{2}}{C_{1}}\,,\\
{\mathcal{A}}^{(2)} & = & N_{c}
\left( \left( \frac{67}{36}-\frac{\pi ^{2}}{12}\right) N_{c}-
\frac{5}{18}N_{f}-\beta _{0}\ln \frac{b_{0}}{C_{1}}\right) \,,
\end{eqnarray}
 where $ N_{f} $ is the number of active quark flavors, $ N_{c}=3 $,
$ C_{F}=4/3 $, $ \beta _{0}=(11N_{c}-2N_{f})/6 $, and 
$ b_{0}\equiv 2e^{-\gamma _{E}}=1.2292... $
. We find that the $ {\mathcal{O}}(\alpha _{S}/\pi ) $ convolution functions
$ {\C}^{(1)}_{g/a}(x) $ can be written as 
\begin{eqnarray}
\C ^{(1)}_{g/g}(x) & = & \delta (1-x)
\left( \frac{\V _{gg\rightarrow \gamma \gamma }(v)}{4}+
\beta _{0}\ln \frac{\mu _{0}}{Q}-\beta _{0}\ln \frac{b_{0}C_{2}}{C_{1}}-
N_{c}\ln ^{2}\frac{b_{0}C_{2}}{C_{1}}\right) \nonumber \\
 &  & -P^{(1)}_{g/g}(x)\ln \frac{\mu _{F}b}{b_{0}},\label{Cgg} \\
\C ^{(1)}_{g/\Sigma }(x) & = & N_{f}C_{F}x-P^{(1)}_{g/\Sigma }(x)
\ln \frac{\mu _{F}b}{b_{0}},\label{Cgq} 
\end{eqnarray}
 where $ P_{g/g}^{(1)}(x) $ and $ P_{g/\Sigma }^{(1)}(x) $ are the
$ {\mathcal{O}}(\alpha _{S}) $ splitting functions. All terms on the right
hand side of Eqs.~(\ref{Cgg}) and (\ref{Cgq}), except for 
$ {\mathcal{V}}_{gg\rightarrow \gamma \gamma }(v) $,
can be obtained from the order-by-order independence of the function 
$ \widetilde{W}(b,Q,v,x_{A},x_{B}) $
on the parameters $ \mu _{0},\, \mu _{F},\, C_{1}, $ and $ C_{2}$,
as well as the universality of the off-diagonal contribution 
$ {\C}^{(1)}_{g/\Sigma }(x) $.
In particular, the term $ \beta _{0}\ln (\mu _{0}/Q) $ occurs because
the LO cross section is $ {\mathcal{O}}(\alpha _{S}^{2}) $, and it implies
that the natural scale for evaluating $ \alpha _{S} $ in Eq.~(\ref{normfact})
is $ \mu _{0}=Q $. The function 
$ {\mathcal{V}}_{gg\rightarrow \gamma \gamma }(v) $
can be obtained from the two-loop corrections to the 
$ gg\rightarrow \gamma \gamma  $
matrix element of Ref.~\cite{BernDeFreitasDixon}. We find 
\begin{equation}
{\mathcal{V}}_{gg\rightarrow \gamma \gamma }=N_{c}\pi ^{2}+
\frac{2{\cal Re}\sum \left[ M^{(1)*}\left( N_{c}F^{L}-N_{c}^{-1}F^{SL}\right) 
\right] }{\sum \left| M^{(1)}\right| ^{2}},
\end{equation}
 where the summation is over the helicities 
$ \lambda _{1},\lambda _{2},\lambda _{3},\lambda _{4} $
of the gluons and photons. The helicity amplitudes 
$ M^{(1)}\equiv M^{(1)}_{\lambda _{1}\lambda _{2}\lambda _{3}\lambda _{4}} $,
$ F^{L}\equiv F^{L}_{\lambda _{1}\lambda _{2}\lambda _{3}\lambda _{4}} $,
and $ F^{SL}\equiv F^{SL}_{\lambda _{1}\lambda _{2}\lambda _{3}\lambda _{4}} $
are given explicitly
in Eqs.~(3.15, 4.7-4.16) of 
Ref.~\cite{BernDeFreitasDixon}.

\begin{figure}[tb]
{\centering \resizebox*{0.49\columnwidth}{!}{\includegraphics{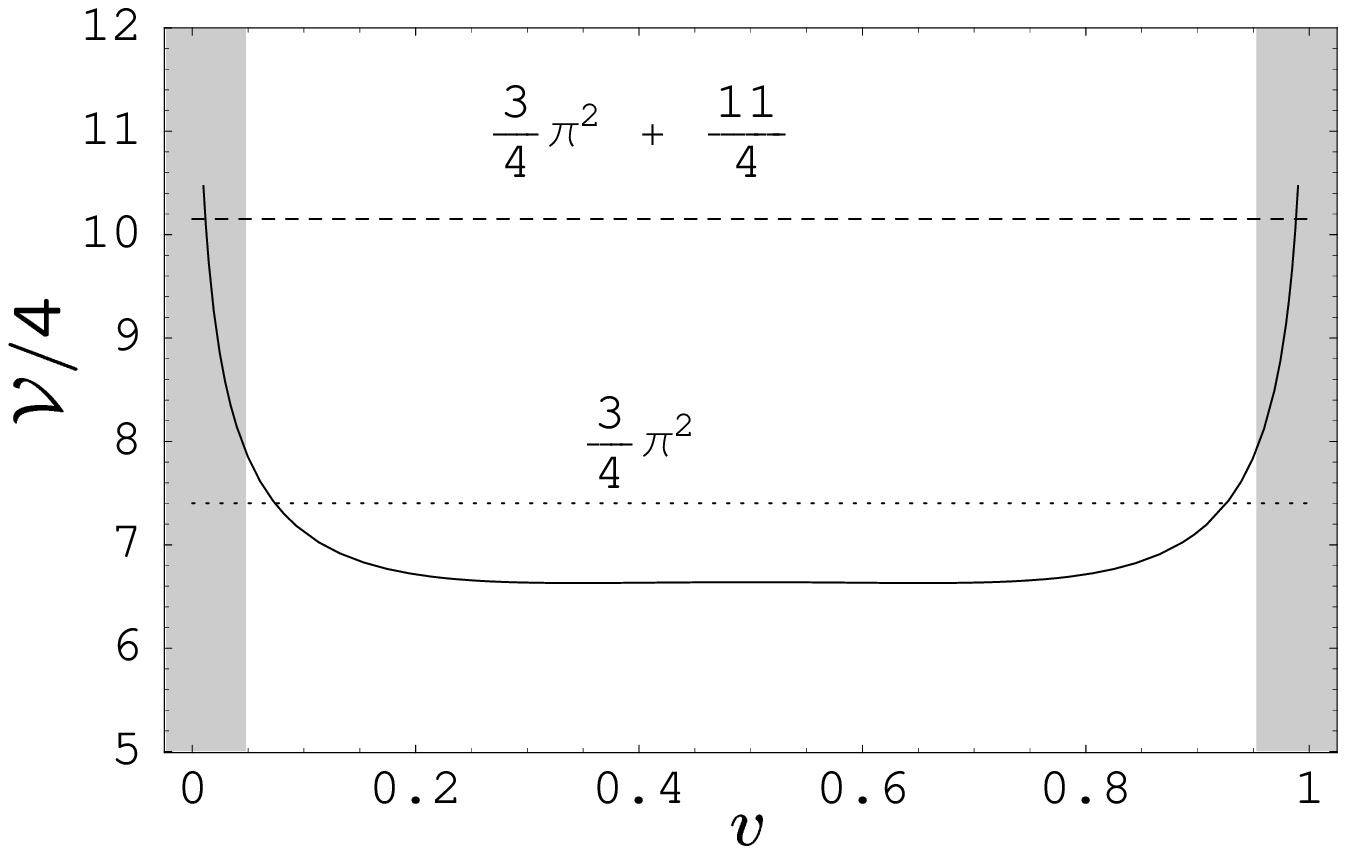}} 
\resizebox*{0.49\columnwidth}{!}{\includegraphics{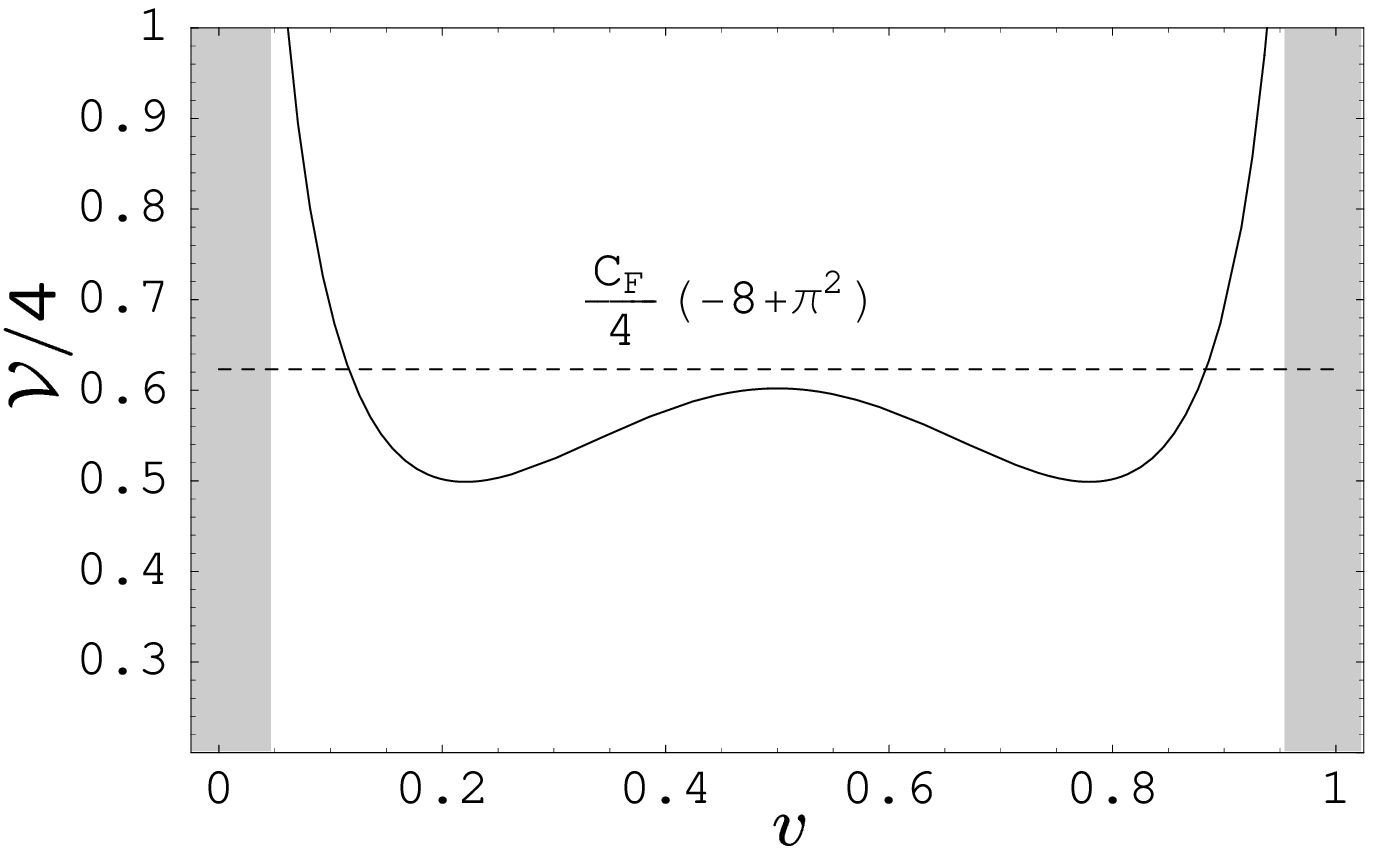}} \par}
{\centering (a)\hspace{2.4in}(b)\par}
\caption{\label{fig:DeltaPiece} Comparison of the functions 
\protect$ {\mathcal{V}}/4\protect \protect $
(a) for \protect$ gg\rightarrow \gamma \gamma \protect \protect $
(solid line) and Higgs boson production \protect$ gg\rightarrow H\protect $
(dashed line); (b) for 
\protect$ q\bar{q}\rightarrow \gamma \gamma \protect \protect $
(solid line) and the Drell-Yan process 
\protect$ q\bar{q}\rightarrow V\protect $
(dashed line). The shaded areas are excluded by the experimental cuts for
\protect$ Q=120\protect \protect $ GeV.}
\end{figure}

In previous studies \cite{BalazsNadolskySchmidtYuan,BalazsBergerMrennaYuan},
before the diphoton two-loop virtual corrections were available, the functions
$ {\C}^{(1)}_{g/a}(x) $ for the process 
$ gg\rightarrow \gamma \gamma X $
were approximated by their counterparts for Higgs boson production, 
$ gg\rightarrow HX $,
calculated in the $ m_{\textrm{top}}\rightarrow \infty  $ limit. The rationale
for this was that both processes are initiated by a $ gg $ initial state
and occur through a quark loop at LO. Thus, the NLO corrections were expected
to be comparable. The functions $ {\C}^{(1)}_{g/a}(x) $ for Higgs
boson production are also given by Eqs.~(\ref{Cgg}) and (\ref{Cgq}), 
except for the replacement of $ \V _{gg\rightarrow \gamma\gamma} $
by~\cite{C1Higgs}
\begin{equation}
\label{nuHiggs}
\V _{gg\rightarrow H}=5N_{c}-3C_{F}+N_{c}\pi ^{2}=11+3\pi ^{2}.
\end{equation}
 Clearly, the use of the Higgs $ {\C} $-functions would
be justified if $ \V _{gg\rightarrow H} $ is numerically close to 
$ \V _{gg\rightarrow \gamma \gamma } $.
To estimate the validity of this approximation, we plot in 
Fig.~\ref{fig:DeltaPiece}(a)
the quantities $ \V _{gg\rightarrow \gamma \gamma }/4 $ and 
$ \V _{gg\rightarrow H}/4 $
as functions of the variable $ v\equiv (1-\nolinebreak \cos \theta ^{*})/2 $.
For the {}``canonical'' choice of parameters $ C_{1}=b_{0}, $ $ C_{2}=1, $
$ \mu _{0}=Q, $ and $ \mu _{F}=b_{0}/b $ we have 
$ {\C}^{(1)}_{g/g}(x)=\delta (1-\nolinebreak x){\mathcal{V}}/4 $;
hence the magnitude of $ {\mathcal{V}} $ completely determines the size
of the $ gg $-initiated NLO correction.

Fig.~\ref{fig:DeltaPiece}(a) shows that 
$ \V _{gg\rightarrow \gamma \gamma }/4 $
is symmetric with respect to $ v\leftrightarrow 1-v $ and becomes singular
in the limits $ v\rightarrow 0 $ and $ v\rightarrow 1 $. These singularities,
which are proportional to powers of $ \ln v $, 
do not contribute to the experimental cross section; they are removed by
cuts on the transverse momenta of the observed photons $ \gamma _{1} $
and $ \gamma _{2} $. For instance, the selection cuts used in 
Ref.~\cite{BalazsNadolskySchmidtYuan}
were $ p_{T}^{\gamma _{1,2}}>25 $ GeV. At LO this imposes the constraint
$ \left( 1-R\right) /2<v<\left( 1+R\right) /2 $, with 
$ R\equiv \left( 1-(2p_{T}^{\gamma _{1}}/Q)^{2}\right) ^{1/2}. $
The excluded regions for $ Q=120 $ GeV are shown by the shaded areas in
Fig.~\ref{fig:DeltaPiece}(a). We see that in most of the allowed
region the function $ \V _{gg\rightarrow \gamma \gamma }/4 $ is nearly
flat, with a numerical value of about 6.65. For comparison, we also plot
in this figure $ \V _{gg\rightarrow H}/4 $, which has a value of 
$ \approx 10.15 $.
Thus, the approximation of substituting the $ {\C}^{(1)}_{g/g}(x) $
coefficient from Higgs production overestimates by about 50\%. On the other
hand, we note that the contribution $ 11/4 $ to $ \V _{gg\rightarrow H}/4 $
comes entirely from the short-distance renormalization to the effective $ Hgg $
operator, which has no counterpart in the $ gg\rightarrow \gamma \gamma  $
process. If we remove this short-distance contribution from
 $ \V _{gg\rightarrow H}/4 $,
we are left with $ N_{c}\pi ^{2}/4\approx 7.40 $, which only overestimates
by about 10\%.

It is interesting to note that the comparable corrections to the process
$ q\bar{q}\rightarrow \gamma \gamma  $ are considerably smaller than for
$ gg\rightarrow \gamma \gamma  $. In Fig.~\ref{fig:DeltaPiece}(b) we
plot the analogous function 
$ {\mathcal{V}}_{q\bar{q}\rightarrow \gamma \gamma }/4 $
(\textit{i.e.,} the coefficient of the $ \delta (1-x) $ term in 
$ {\C}^{(1)}_{q/q}(x) $),
which was given in Refs.~\cite{BalazsBergerMrennaYuan,BaileyOwensOhnemus}.
We see that it is equal to $ 0.5-0.6 $ in most of the kinematical region
selected by the LHC cuts, which is much less than the value of 6.65 that
we found for the $ gg $-initiated process. In this figure we also plot
the analogous coefficient $ {\mathcal{V}}_{DY}/4 $ for the Drell-Yan process, 
which differs from $ {\mathcal{V}}_{q\bar{q}\rightarrow \gamma \gamma }/4 $
by less than $ 5-20\% $ over this kinematic range.

Since the function $ \V _{gg\rightarrow \gamma \gamma } $ corrects only
the $ \delta (1-x) $ piece of $ {\C}^{(1)}_{g/g}(x) $, and
it does not depend on the impact parameter $ b$, its primary effect is
to change the overall normalization of the transverse momentum distribution,
but not its shape. In Ref.~\cite{BalazsNadolskySchmidtYuan}, a K-factor
was defined as the ratio of the NLO resummed cross section to the 
LO non-resummed
cross section, using the corresponding PDFs in the numerator and denominator.
By approximating the function $ \V _{gg\rightarrow \gamma \gamma } $ by
the analogous one for Higgs production, Eq.~(\ref{nuHiggs}), the K-factor
for the process $ gg\rightarrow \gamma \gamma  $ was estimated to be 1.45-1.75.
We can now consider the impact of the correct function on the K-factor. Given
that the contribution of 
$ ({\C}_{g/\Sigma }\otimes f_{\Sigma })(x, b;\mu _{F}) $
constitutes less than $ 25\% $ of the contribution of 
$ ({\C}_{g/g}\otimes f_{g})(x, b;\mu _{F}) $
in the central rapidity region, we estimate the correct 
$ gg\rightarrow \gamma \gamma X $
K-factor to be about 1.2-1.5. 
Furthermore, we can use Fig.~2 in 
Ref.~\cite{BalazsNadolskySchmidtYuan}
to find the corrected K-factor for all included subprocesses to be about
1.3 at $ Q=80 $ GeV and 1.6 at $ Q=150 $ GeV. 
We note that the resummed K-factors for the $ gg\rightarrow \gamma \gamma  $
subprocess
are slightly different than the fixed-order K-factors obtainable from
Fig.~4(a) in 
Ref.~\cite{BernDixonSchmidt};  however, this difference is primarily
due to the fact that the renormalization scale was chosen to be $\mu_0=Q/2$
in Ref.~\cite{BernDixonSchmidt} and that the different selection cuts
used in that paper produced a kinematic enhancement of the K-factor
for $Q$ near 80 GeV.
Of course, these first estimates of the corrected resummed K-factors can
be further refined by repeating a detailed Monte-Carlo study as in 
Ref.~\cite{BalazsNadolskySchmidtYuan}.

\begin{figure}[tb]
{\centering \resizebox*{0.55\columnwidth}{!}
{\includegraphics{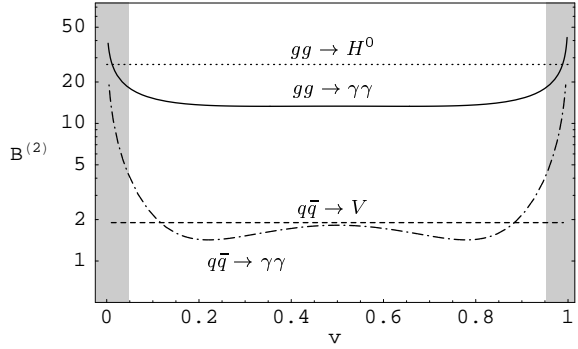}} \par}
\caption{\label{fig:B2} Comparison of the coefficient 
\protect$ {\mathcal{B}}^{(2)}\protect \protect $
in various particle reactions. The shaded areas are excluded 
by the experimental
cuts for \protect$ Q=120\protect \protect $ GeV.}
\end{figure}

Recently, it has been shown that the remaining 
$ {\mathcal{O}}(\alpha _{S}^{2}/\pi ^{2}) $
coefficient in the Sudakov factor, $ {\mathcal{B}}^{(2)} $, can also be
obtained from the NLO cross section, using the universality 
of the real emission
corrections and the general structure of the virtual corrections in the soft
and collinear limits \cite{dFG}. Following this argument, we obtain 
\begin{eqnarray}
{\mathcal{B}}_{gg\rightarrow X}^{(2)} & = & -\frac{\delta P_{g/g}^{(2)}}{2}+
\beta _{0}\left( \frac{\V _{gg\rightarrow X}}{4}+
\frac{N_{c}\pi ^{2}}{12}\right) +\beta _{0}^{2}
\ln \frac{\mu _{0}}{Q}\nonumber \\
 &  & -2N_{c}\left( \left( \frac{67}{36}-\frac{\pi ^{2}}{12}\right) N_{c}-
\frac{5}{18}N_{f}\right) \ln \frac{b_{0}C_{2}}{C_{1}}\nonumber \\
 &  & +\beta _{0}N_{c}\left( \left( \ln \frac{b_{0}}{C_{1}}\right) ^{2}-
\left( \ln C_{2}\right) ^{2}\right) -\beta _{0}^{2}\ln C_{2},\label{b2gg} 
\end{eqnarray}
 which is valid both for Higgs boson production and diphoton production.
In this formula the $ \delta (1-x) $ part of the $ {\cal O}(\alpha _{S}^{2}) $
splitting function for $ g\rightarrow g $ is 
\begin{equation}
\delta P^{(2)}_{g/g}\, \, =\, \, N_{c}^{2}\left( \frac{8}{3}+
3\zeta (3)\right) -\frac{1}{2}N_{f}C_{F}-\frac{2}{3}N_{f}N_{c}\, \, ,
\end{equation}
 where $ \zeta (n) $ is the Riemann zeta function, with 
$ \zeta (3)=1.202057\dots  $.
For Higgs boson production, Eq.~(\ref{b2gg}) has been corroborated by direct
calculation from the NLO transverse momentum 
distributions~\cite{glosserschmidt}.
In Fig.~\ref{fig:B2} we plot the $ {\cal B}^{(2)} $ coefficient functions
for various processes, with the canonical choice of parameters and $ N_{f}=5 $.
From this plot, we see that $ {\mathcal{B}}^{(2)}_{gg\rightarrow H} $
is almost exactly twice as large as 
$ {\mathcal{B}}^{(2)}_{gg\rightarrow \gamma \gamma } $
over most of the allowed kinematic region, and both coefficients are 
considerably
larger than those for the $ q\bar{q} $-initiated processes.

In the standard CSS formalism, the functions $ {\mathcal{B}} $ and 
$ {\C}_{g/a} $
are process-dependent, as seen explicitly above. Ref.~\cite{CdFG} proposed
a modified resummation formula, which removes from these functions all terms
associated with hard virtual QCD corrections to the LO process. Such hard
corrections are absorbed in a new function $ {\mathcal{H}}(\alpha _{S}) $,
so that the alternate formula for $ \widetilde{W}(b,Q,v,x_{A},x_{B}) $
is 
\begin{eqnarray}
&& \widetilde{W}(b,Q,v,x_{A},x_{B})=\frac{\sigma _{0}(\alpha _{S}
(\mu _{0}))}{S}{\mathcal{H}}(\alpha _{S}(\mu _{0})) \nonumber \\
 && \times \exp \left\{ -\int _{C_{1}^{2}/b^{2}}^{C^{2}_{2}Q^{2}}
\frac{d\bar{\mu }^{2}}{\bar{\mu }^{2}}\Bigl [{\mathcal{A}}(\alpha _{S}(
\bar{\mu }))\ln \frac{C_{2}^{2}Q^{2}}{\bar{\mu }^{2}}+
{\mathcal{B}}^{\prime }(\alpha _{S}(\bar{\mu }))\Bigr ]\right\} \nonumber \\
 && \times \sum _{a,b}\left[ \C ^{\prime }_{g/a} \otimes f_{a}
\right] (x_{A}, b;\mu _{F}, \alpha _{S}(\mu _{F}))
\left[ \C ^{\prime }_{g/b} \otimes f_{b}\right] 
(x_{B}, b;\mu _{F}, \alpha _{S}(\mu _{F})). \label{altCSS}
\end{eqnarray}
Here we can expand $ {\mathcal{B}}^{\prime }(\alpha _{S}) $ and 
$ {\C}^{\prime }_{g/a}(x, b; \mu_F, \alpha _{S}) $
as a series in $ \alpha _{S} $ exactly as the functions 
$ {\mathcal{B}}(\alpha _{S}) $
and $ {\C}_{g/a}(x, b; \mu_F, \alpha _{S}) $, and the function 
$ {\mathcal{H}}(\alpha _{S}) $
can similarly be expanded as 
${\mathcal{H}}(\alpha _{S})\, \, =\, \, 1+\sum _{n=1}^{\infty }
\left( {\alpha _{S}}/{\pi }\right) ^{n}{\mathcal{H}}^{(n)}$.

In this formulation, there is a {}``scheme-dependent'' ambiguity in the
definition of $ {\C}^{\prime }_{g/g} $, $ {\mathcal{B}}^{\prime } $,
and $ {\mathcal{H}} $, since a change in $ {\mathcal{H}} $ can be compensated
by redefinitions of $ {\C}^{\prime }_{g/g} $ and 
$ {\mathcal{B}}^{\prime } $.
A reasonable choice of scheme is to define 
\begin{equation}
{\mathcal{H}}^{(1)}_{gg\rightarrow X}\, \, =\, \, 
\frac{\V _{gg\rightarrow X}}{2}+2\beta _{0}\ln \frac{\mu _{0}}{Q},
\end{equation}
 so that $ \C ^{(1)\prime }_{g/g}(x) $ vanishes for the canonical 
choice of parameters.
In this scheme, which is similar to the
`NS resummation scheme' of Ref.~\cite{CdFG}, we obtain 
\begin{eqnarray}
\C ^{(1)\prime }_{g/g}(x) & = & \delta (1-x)\left( -\beta _{0}
\ln \frac{b_{0}C_{2}}{C_{1}}-N_{c}\ln ^{2}\frac{b_{0}C_{2}}{C_{1}}\right) -
P^{(1)}_{g/g}(x)\ln \frac{\mu _{F}b}{b_{0}}\, \, ,\\
{\mathcal{B}}_{gg\rightarrow X}^{(2)\prime } & = & -
\frac{\delta P_{g/g}^{(2)}}{2}+\beta _{0}N_{c}\frac{\pi ^{2}}{12}+
\beta _{0}N_{c}\left( \left( \ln \frac{b_{0}}{C_{1}}\right) ^{2}-
\left( \ln C_{2}\right) ^{2}\right) \nonumber \\
 &  & -2N_{c}\left( \left( \frac{67}{36}-\frac{\pi ^{2}}{12}\right) N_{c}-
\frac{5}{18}N_{f}\right) \ln \frac{b_{0}C_{2}}{C_{1}}-
\beta _{0}^{2}\ln C_{2}\, \, ,
\end{eqnarray}
 as well as $ {\mathcal{B}}^{(1)'}={\mathcal{B}}^{(1)} $ and 
$ {\C}^{(1)\prime }_{g/\Sigma }(x)=
{\C}^{(1)}_{g/\Sigma }(x) $.
The advantage of this formulation for diphoton production is that it allows
us to shift all dependence on the kinematical variable $ v $ from 
$ {\C}^{\prime }_{g/g} $
and $ {\mathcal{B}}^{\prime } $ into the single hard factor $ {\mathcal{H}} $.
This choice makes sense physically, since this kinematical dependence is
a property of the hard $ gg\rightarrow \gamma \gamma  $ process, rather
than of soft or collinear effects. This formulation also makes more obvious
the fact that the function $ \V _{gg\rightarrow \gamma \gamma } $ affects
the normalization, but not the shape, of the transverse momentum distribution.
A similar modification can be made to the 
$ q\bar{q}\rightarrow \gamma \gamma X $
resummation formula.

In conclusion, we have calculated the remaining unknown parts at 
$ {\mathcal{O}}(\alpha _{S}/\pi ) $
in the resummed cross section for the production of photon pairs in gluon-gluon
fusion at small $ Q_{T} $. We found that the approximation of the function
$ {\C}_{g/g}(x, b; \mu_F) $ in the process 
$ gg\rightarrow \gamma \gamma  $
by its counterpart from Higgs boson production overestimates the 
$ gg\rightarrow \gamma \gamma X $
resummed K-factor by about $ 15-20\% $, and it overestimates the K-factor
for the total diphoton production process by about $ 5-10\% $. We have
also calculated the $ {\mathcal{O}}(\alpha _{S}^{2}/\pi ^{2}) $ coefficient
$ {\mathcal{B}}^{(2)} $ in the perturbative Sudakov factor. We predict
that the impact of the coefficient $ {\mathcal{B}}^{(2)} $ on the shape
of transverse momentum distributions in gluon fusion is more substantial
then in the process $ q\bar{q}\rightarrow \gamma \gamma  $, and that it
will improve the matching of the resummed calculation with the fixed-order 
calculation at intermediate~$ Q_{T} $ .

\end{document}